\documentclass{INTERSPEECH2023}

\interspeechcameraready

\usepackage{amsmath,graphicx}
\usepackage{tabularx} 
\usepackage{booktabs} 
\usepackage{multirow}
\usepackage{multicol}
\usepackage{hyperref}
\usepackage[table]{xcolor}
\usepackage{lipsum}

\newcommand\blfootnote[1]{%
  \begingroup
  \renewcommand\thefootnote{}\footnote{#1}%
  \addtocounter{footnote}{-1}%
  \endgroup
}

\title{Model Extraction Attack against Self-supervised Speech Models}
\name{Tsu-Yuan Hsu$^{1*}$, Chen-An Li$^{2*}$, Tung-Yu Wu$^3$, Hung-yi Lee$^4$}
\address{
  $^1$$^2$$^3$$^4$ College of Electrical Engineering and Computer Science, National Taiwan University}
\email{\{b08201047,b08902123,b08901133,hungyilee\}@ntu.edu.tw}

\begin{document}

\maketitle
\begin{abstract}
Self-supervised learning (SSL) speech models generate meaningful representations of given clips and achieve incredible performance across various downstream tasks. Companies can provide services of these models by building APIs. However, each of these APIs may suffer a model extraction attack (MEA), which refers to an adversary stealing the model functionality with limited query access. In this work, we propose an MEA 
 framework against the SSL speech models. Our MEA framework learns multiple output representations of given clips to extract the target SSL speech models. We demonstrate various selection methods on speech corpus to construct limited query access. We also study the MEA on different speech corpus. We evaluate the effectiveness of our MEA framework on four diverse downstream tasks. To our knowledge, this is the first attempt to steal a large-scale speech model.
\end{abstract}

\blfootnote{*equal contributions}
\section{Introduction}
\label{sec:intro}
Recent advances in self-supervised learning (SSL) speech models \cite{baevski2020wav2vec, hsu2021hubert, chen2022wavlm} build meaningful representations of speech and achieve incredible performance in many tasks \cite{yang21c_interspeech}. Regarding the current SSL-based natural language processing APIs, such as official GPT-3 \cite{brown2020language} APIs, which provide services for generating data embeddings, it can be expected that SSL speech-processing APIs would also come into sight in the future. This kind of APIs take text or speech provided by users as input and generate corresponding feature representations for the training of downstream models. 

However, each of these APIs may suffer a model extraction attack (MEA), which refers to an adversary stealing the model functionality by limited query access. MEA has posed a non-negligible threat to online deep learning applications. Previous work \cite{tramer2016stealing} has shown that the adversary may extract models used in remote APIs simply by querying them. Since the training of models and the collection of datasets may have cost a tremendous amount of time and money, this kind of attack causes sizeable financial losses to victimized companies. Hence, it is an urgent task for researchers to study how the adversary may perform the attack.

MEAs against different classes of machine learning and deep learning models are broadly studied. Through some direct queries to the remote API, simple regression models and multilayer perceptrons (MLPs) can be easily stolen \cite{tramer2016stealing}. Convolutional neural networks (CNNs) are also vulnerable to the attack \cite{correia2018copycat, orekondy2019knockoff}. \cite{correia2018copycat} randomly selects images on hand, queries the API with selected images to fetch fake labels, and utilizes the image-label pairs to train the local surrogate model. Knockoff Nets \cite{orekondy2019knockoff} adopts reinforcement learning to actively sample images, which surpasses the random-selection approach. Recurrent Neural Networks (RNNs) have also been studied to be attacked. \cite{takemura2020model} studies on attacks based on features of RNNs and LSTMs for classification and regression tasks. Advanced models such as graph neural networks (GNNs) have also been examined \cite{shen2022model, wu2022model}. For instance, \cite{shen2022model} demonstrates that, after collecting query graphs, a surrogate model can be effectively learned by minimizing RMSE loss between the remote GNN-based API's and its graph responses. Besides model-type-specific extraction attacks, there are also some works \cite{wang2018stealing, oh2019towards} committed to stealing certain information of remote APIs. Model hyperparameters have been pointed out as a potential target, and a framework is proposed to verify the feasibility of hyperparameter extraction \cite{wang2018stealing}. Metamodel methods \cite{oh2019towards} learn a classifier to predict model attributes, such as model architectures, adopted optimizers, and types of training datasets.

Large-scale SSL models \cite{baevski2020wav2vec, hsu2021hubert, chen2022wavlm, devlin-etal-2019-bert, yang2019xlnet, liu2019roberta} are more critical potential targets of MEA since the SSL-model-based APIs can serve as a powerful feature extractor to generate representations of input data that help users implement various applications such as text question answering (QA) with BERT \cite{devlin-etal-2019-bert} and automatic speech recognition (ASR) with wav2vec 2.0 \cite{baevski2020wav2vec}. The training and fine-tuning of an SSL model require much time and effort. However, the MEAs against SSL models are still underexplored. Though there have been some works \cite{Krishna2020Thieves, chen2021killing, he-etal-2021-model, zanella2021grey} investigating MEA against text models, to our knowledge, approaches for speech models have not been discussed. Moreover, current text-model-based methods still exist some restrictions. \cite{Krishna2020Thieves, chen2021killing, he-etal-2021-model} apply random sampling strategy to select query data, while \cite{zanella2021grey} presents a hybrid strategy that integrates algebraic-based and learning-based methods. Nonetheless, they all assume attackers have access to output logits of downstream tasks and that each time the attack merely focuses on stealing one task. 
\blfootnote{Note that this work aims to capture the research community's attention to potential issues in terms of speech-based SSL APIs, rather than simply an attempt to conduct MEA. With this work, we anticipate more discussions in this field to help build a more robust and comprehensive ecosystem of speech-based Machine Learning as a Service (MLaaS).}

In this work, we propose and implement the MEA against SSL speech models. In particular, we design several clip-selection methods that identify informative data. The selected clips are further used as queries to the victim model to get corresponding representations for the local surrogate model's supervised training. This enables our surrogate model to approximate the victim model's performances with only a small number of clips, i.e., queries. We demonstrate MEA on various selection methods and different speech corpus. Specifically, four diverse downstream tasks are conducted in Section \ref{sec:exp} to evaluate the effectiveness of our proposed clip selection methods and the whole extraction pipeline. To our knowledge, this is the first attempt to steal a large-scale speech model. Furthermore, since our framework's presented active-sampling methods only need access to data embeddings instead of logits, we can extract the remote SSL speech model directly rather than just a downstream task. 

\begin{figure}[htb]
  \resizebox{\linewidth}{!}{\centerline{\includegraphics[width=9cm]{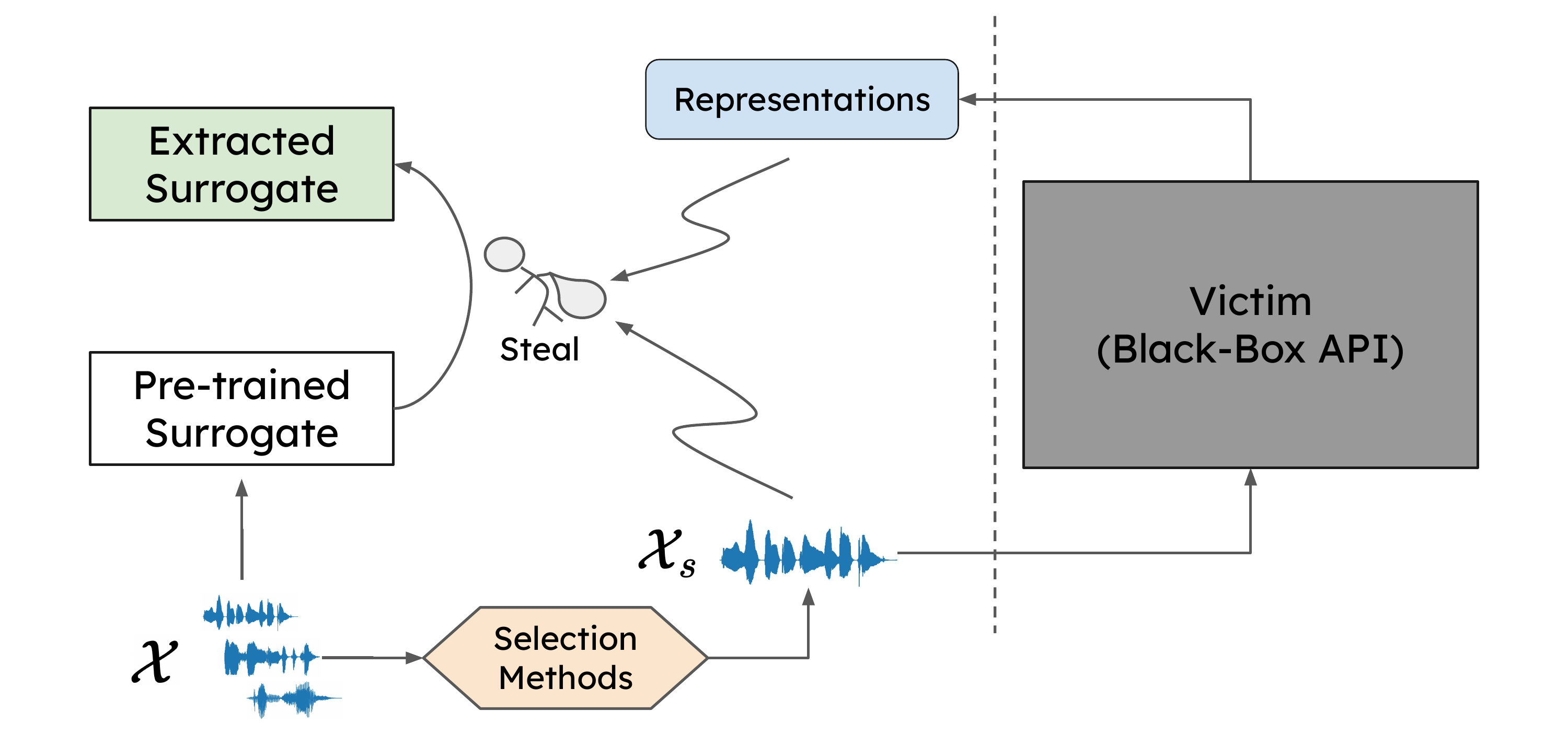}}}
  \caption{Illustration of our model extraction attack framework.}
  \label{fig:framework}
\end{figure}

\section{Methods}
\label{sec:methods}

In this paper, the victim model refers to the model being queried, such as the APIs mentioned in Section \ref{sec:intro}. The pre-trained surrogate model refers to the model pre-trained on unlabeled corpus and the extracted surrogate model refers to the model after performing model extraction on the victim model. We assume the requests to the victim model are limited, and the knowledge of the victim model's architecture and its training corpus is lacking. The query limitation is defined as the total length of waveforms. The information returned by the victim model includes representations of input data.

As shown in Figure \ref{fig:framework}, our model extraction attack process includes several steps: 
\begin{enumerate}[label=(\arabic*), leftmargin=1.5\parindent]
    \item Pre-train a model on unlabeled corpus $\mathcal{X}$ to get the \textit{Pre-trained Surrogate}. 
    \item An active selection method is applied to sample a small portion of clips from dataset $\mathcal{X}$ to form a subset $\mathcal{X_S}$. The sampling process is done until the total length of the waveforms in $\mathcal{X_S}$ is no less than the preset length limitation $H$. 
    \item The \textit{Pre-trained Surrogate} is trained with $\mathcal{X_S}$ and the obtained \textit{Victim}'s representations to perform the model extraction, i.e., steal the functionality of the \textit{Victim} to get the \textit{Extracted Surrogate}.
\end{enumerate}

\subsection{Selection Methods}
\label{sec:method:sm} 
In this section, we elaborate on several proposed clip selection methods used to construct $\mathcal{X_S}$ for model extraction.

\subsubsection{SSL Pre-training Loss Selection} 
\label{sec:method:pre} 

We state that SSL pre-training loss can serve as a good metric to sample waveforms. Clips with high pre-training loss are regarded as hard samples which baffle the current surrogate model and are thus worth engaging in the afterward teacher-student supervised training. As a result, we evaluate the dataset $\mathcal{X}$ on the pre-trained surrogate model and calculate each waveform's pre-training loss. The dataset $\mathcal{X_S}$ iteratively samples waveform with the highest loss in $\mathcal{X}$ until its total clip length reaches $H$.

\subsubsection{Content-based Selection}
\label{sec:method:sm:cb}
A selection approach based on the acoustic content of speech clips is also proposed. We argue that, with the knowledge of each recording's acoustic information, it is possible to sample a small portion of clips that represents the overall distribution of the whole corpus. To achieve this, we first leverage the pre-trained surrogate model to generate all the waveform representations. Secondly, each waveform timestamp's token is determined by feeding its representation into a clustering model fit on 10\% of the representations, with consecutive same class removed. We then take these tokens as the content of a speech clip. Finally, we iteratively sample the corpus with the farthest point sampling (FPS) \cite{eldar1997farthest}. To be specific, after randomly picking the first clip, we calculate its token-based trigram Jaccard distance to all other clips:
\begin{align}
    D_{x, y} = 1 - J(X, Y) = 1 - \frac{|X\cap Y|}{|X\cup Y|},
\end{align}
where $x$ and $y$ are two distinct clips (a sampled clip and an unsampled one in our case), while $X$ and $Y$ are their token trigram sets. Specifically, each element in the set is a set containing three consecutive tokens generated by the clustering model. $J(X, Y)$ is the Jaccard similarity. Among all unselected clips, we choose the one with the farthest distance (lowest similarity) to the previously-selected clip as the second sampled clip. If there have been multiple clips already sampled, an unsampled clip's distance to them is defined as the Jaccard distance to its nearest sampled clip. The FPS process is repeated until the total length of sampled clips is no less than $H$.

\subsubsection{Transcription-based Selection}
\label{sec:method:sm:tb}
In this method, we select clips with different content. We assume that every speech in the dataset has its corresponding transcription. We utilize a pre-trained language model to generate \texttt{[CLS]} token embeddings for speech transcriptions. Next, a clustering model is fit on \texttt{[CLS]} token embeddings to get clustering labels. Finally, we evenly select the corresponding clips in $\mathcal{X}$ from each cluster to increase the data diversity into $\mathcal{X}_s$.

\begin{table*}[t]
    \caption{Evaluation results with querying dataset LibriSpeech 960-hour for KS, IC, and ER in accuracy and SD in diarization error rate. The first three rows are w/o any query. Baseline$_R$ refers to performing model extraction on Random Surrogate w/ random selection. Baseline$_P$, Transcription, SSL, and Content refer to the Extracted Surrogate w/ random selection and the selection methods elaborated in Section 2. Topline$_H$ and Topline$_W$ refer to the Extracted Surrogate w/ unlimited queries to the HuBERT Base and WavLM Base+, respectively. Best result among Baseline$_R$, Baseline$_P$, Transcription (if any), SSL, and Content of each experiment is marked in bold.
    }
    \centering
    \resizebox{\textwidth}{!}{
    \begin{tabular}{l c | c c c | c c c | c c c | c c c}
        \specialrule{2.5pt}{1pt}{1pt}
         \rowcolor{gray!20}&& \multicolumn{3}{c|}{\textbf{KS (Acc $\uparrow$)}} & \multicolumn{3}{c|}{\textbf{IC (Acc $\uparrow$)}} & \multicolumn{3}{c|}{\textbf{ER (Acc $\uparrow$)}} & \multicolumn{3}{c}{\textbf{SD (DER $\downarrow$)}} \\
        \cmidrule{1-14}
        \multicolumn{2}{l|}{\textit{Pre-trained Surrogate}} & \multicolumn{3}{c|}{92.41} & \multicolumn{3}{c|}{77.91} & \multicolumn{3}{c|}{57.7} & \multicolumn{3}{c}{7.86} \\
        \midrule
        \multicolumn{2}{l|}{\textit{HuBERT Base}} & \multicolumn{3}{c|}{96.30} & \multicolumn{3}{c|}{98.34} & \multicolumn{3}{c|}{64.92} & \multicolumn{3}{c}{5.88} \\
        \midrule
        \multicolumn{2}{l|}{\textit{WavLM Base+}} & \multicolumn{3}{c|}{97.37} & \multicolumn{3}{c|}{99.00} & \multicolumn{3}{c|}{68.65} & \multicolumn{3}{c}{3.50} \\
        \midrule
        \multicolumn{2}{l|}{\textit{Topline}$_{H}$} & \multicolumn{3}{c|}{95.85} & \multicolumn{3}{c|}{94.83} & \multicolumn{3}{c|}{62.74} & \multicolumn{3}{c}{6.76} \\
        \midrule
        \multicolumn{2}{l|}{\textit{Topline}$_{W}$} & \multicolumn{3}{c|}{96.33} & \multicolumn{3}{c|}{95.68} & \multicolumn{3}{c|}{64.02} & \multicolumn{3}{c}{6.72} \\
        \specialrule{2.5pt}{1pt}{1pt}

        \rowcolor{gray!20}\bf\textit{Method} & \bf\textit{Victim Model} & \textit{0.1h} & \textit{1h} & \textit{10h} & \textit{0.1h} & \textit{1h} & \textit{10h} & \textit{0.1h} & \textit{1h} & \textit{10h} & \textit{0.1h} & \textit{1h} & \textit{10h} \\
        \specialrule{1.5pt}{1pt}{1pt}
        \textit{Baseline$_R$} & HuBERT Base & 83.47 & 89.65 & 94.26 & 63.67 & 79.73 & 87.29 & 56.41 & 57.72 & 60.39 & 11.02 & 10.05 & 7.56 \\
        \midrule
        \textit{Baseline$_P$} & HuBERT Base & 92.05 & 94.16 & 94.35 & 75.43 & 87.43 & 88.98 & 56.91 & 60.28 & 60.06 & 9.15 & 7.64 & 7.18 \\
        \midrule
        \textit{Transcription} & HuBERT Base & 92.11 & 93.57 & 94.22 & 76.19 & 84.29 & 90.17 & 58.68 & 59.78 & 61.48 & 9.61 & \textbf{7.63} & 7.61 \\
        \midrule
        \textit{SSL} & HuBERT Base & \textbf{93.12} & \textbf{95.30} & 95.07 & \textbf{79.09} & \textbf{88.74} & 90.43 & \textbf{59.34} & 60.03 & \textbf{62.11} & \textbf{9.14} & 8.32 & 7.16 \\
        \midrule
        \textit{Content} & HuBERT Base & 92.08 & 94.45 & \textbf{95.29} & 77.54 & 88.72 & \textbf{91.27} & 57.99 & \textbf{60.36} & 61.49 & 9.28 & 7.80 & \textbf{6.93}\\
        
        \specialrule{1.75pt}{1pt}{1pt}
        \textit{Baseline$_R$} & WavLM Base+ & 81.50 & 88.93 & 93.02 & 60.77 & 78.01 & 86.45 & 55.18 & 58.00 & 61.55 & 12.46 & 10.48 & 7.56 \\
        \midrule
        \textit{Baseline$_P$} & WavLM Base+ & 93.22 & 93.51 & 94.35 & 78.30 & 87.77 & 91.48 & 59.41 & 61.56 & 61.95 & 9.62 & \textbf{7.31} & 7.52 \\
        \midrule
        \textit{SSL} & WavLM Base+ & \textbf{93.67} & \textbf{94.68} & \textbf{95.46} & \textbf{82.92} & 90.11 & \textbf{92.33} & \textbf{59.87} & 61.78 & \textbf{63.71} & 8.96 & 7.74 & 7.23 \\
        \midrule
        \textit{Content} & WavLM Base+ & 93.18 & 94.55 & 95.13 & 79.41 & \textbf{90.4} & 92.17 & 58.84 & \textbf{62.29} & 62.70 & \textbf{8.74} & 7.77 & \textbf{6.84} \\
        \specialrule{2.5pt}{1pt}{1pt}
    \end{tabular}
    }
    \label{table:lsds}
\end{table*}
\subsection{Model Extraction} 
\label{sec:method:me}
In this paper, we adopt the objective function of DistilHuBERT \cite{chang2022distilhubert} to perform model extraction. We assume that the victim model returns multiple representations per query because weighted-sum multiple hidden states of the pre-train model could significantly improve the downstream performance \cite{yang21c_interspeech}. The surrogate model is followed by multiple separate prediction heads which learn the victim model's representations from different layers. 

Given victim model's $n$-th output representation $h^{(n)}$ and the prediction head vector learn from victim model’s $n$-th output $\hat{h}^{(n)}$, The objection function $\mathcal{L}_{\text{extract}}$ can be shown as follows:
\begin{align}
    \mathcal{L}_{\text{extract}}^{(n)} &= 
    \mathcal{L}_{cos}^{(n)} + \mathcal{L}_{l1}^{(n)} \nonumber \\ 
        &= \sum_{t=1}^{T}
        \left[-
        \log \sigma 
            \left(\cos\left(h_t^{(n)}, \hat{h}_t^{(n)}\right)\right) + \frac{1}{D} \lVert h_t^{(n)} - \hat{h}_t^{(n)} \rVert_1
        \right]
\end{align}
\begin{align}
    \mathcal{L}_{\text{extract}} &= \sum_{n \in N} \mathcal{L}_{\text{extract}}^{(n)} = \sum_{n \in N} \left[ 
    \mathcal{L}_{cos}^{(n)} + \mathcal{L}_{l1}^{(n)} \right] 
\end{align}
where $t \in [T]$, $T$ is the number of timestamps, $N$ is the set of the layers' indices, $\sigma(\cdot)$ denotes sigmoid function, $cos(\cdot, \cdot)$ denotes cosine similarity function, $D$ is the feature dimension of the representation.

\section{Experiments}
\label{sec:exp}
\subsection{Experimental Setup} 
\label{sec:exp:setup}
Experiments are implemented with s3prl v0.3.4 and fairseq \cite{ott2019fairseq} 0.12.2. We use fairseq for pre-training our surrogate model. For s3prl, we use it for performing model extraction and evaluation.

\subsubsection{Data and Preprocessing} 
\label{sec:exp:setup:data} 
We use LibriSpeech 960-hour \cite{panayotov2015librispeech} to pre-train the surrogate model and as the querying dataset to the victim model in our main results. Extensive experiments are conducted using Wall Street Journal \cite{paul1992design} and Aishell-1 \cite{bu2017aishell} as querying datasets. Considering the victim model should have an instance-wise querying-length constraint, the maximum input sequence length is set to 15.6 seconds, and each clip longer than 15.6s is split. Also, we set the minimum waveform length to 2s, which means the clip less than 2 seconds will be dropped.

\subsubsection{Victim Model}
\label{sec:exp:setup:tm} 
We consider HuBERT Base and WavLM Base+ as the victim models.
Each of them has a 12-layer transformer encoder. Features from different layers contain various information \cite{chen2022wavlm,chang2022distilhubert}, such as speaker-related, semantic, and content information.
Therefore, we assume the victim model returns the representations of each transformer layer, and the 4$^{\text{th}}$, 8$^{\text{th}}$, and 12$^{\text{th}}$-layer representations are used to perform the model extraction.

The pre-training dataset of HuBERT Base is exactly the same as our querying dataset, i.e. LibriSpeech 960-hour while that of WavLM Base+ is a 94k-hour dataset, implying the overlap with our querying dataset is less than 1.1\%. We assume the latter one to be more realistic. That is, the API (victim model) is pre-trained on a large-scale dataset and should cope with speech clips from various domains well. However, we only have a small piece of the dataset, and we intend to query the victim model as efficiently as possible.

\subsubsection{Surrogate Model}
\label{sec:exp:setup:lm}

Our surrogate model is initialized as a 7-layer CNN extractor following a 2-layer transformer encoder, called \textit{Random Surrogate}. HuBERT pre-training is applied on \textit{Random Surrogate} with LibriSpeech 960-hour to obtain \textit{Pre-trained Surrogate}. The first iteration is trained for 250k steps by the MFCC clustered labels. The second iteration is trained for 400k steps by the clustered labels generated from the $1^{\text{st}}$ transformer encoder layer features of first-iteration \textit{Pre-trained Surrogate}.

In the selection stage, the corresponding features of sampled clips are retrieved by querying the victim model, where the clips are obtained by either random, pre-training loss, content-based, or transcription-based selection. We adopt the k-means model \cite{lloyd1982least} as the clustering model used for content- and transcription-based selection where the number of clusters is 250. For SSL pre-training loss selection, we use the self-supervised cluster-prediction loss of HuBERT. For transcription-based selection, we use RoBERTa \cite{liu2019roberta} as our language model.

In the model extraction stage, our surrogate model (\textit{Random Surrogate} or \textit{Pre-trained Surrogate}) is trained on the clip-feature pairs for $10k\times H$ steps with batch size 24, where $H$ is the hour of sampled clips and $H=0.1,1,10$ are considered in this paper. The learning rate linearly increases to 0.0002 in the first 7\% of the total steps and then linearly decreases to zero in the left steps.

\subsection{Evaluation}
We use SUPERB \cite{yang21c_interspeech} benchmark to evaluate our surrogate model. In SUPERB, the surrogate model is frozen and a trainable linear layer is adopted to weigh each layer's output features of our surrogate model, including one CNN extractor's, two transformer encoder layers', and three prediction heads' outputs. The downstream models then use the weighted feature as the input. In this paper, four downstream tasks, including keyword spotting (KS) \cite{speechcommands}, intent classification (IC) \cite{Lugosch2019}, emotion recognition (ER) \cite{busso2008iemocap}, and speaker diarization (SD) \cite{cosentino2020librimix}, are selected to evaluate the surrogate model.

\begin{table}
    \centering
    \caption{Performances of SID and SD with victim model WavLM Base+. Results better than \textit{Pre-trained Surrogate} (65.80 for SID and 7.86 for SD) are marked in bold.}
    \resizebox{\linewidth}{!}{
    \begin{tabular}{l | c c c | c c c}
      \toprule
      & \multicolumn{3}{c|}{\textbf{SID}} & \multicolumn{3}{c}{\textbf{SD}} \\
      \cmidrule{2-7}
      \textit{Method} & \textit{0.1h} & \textit{1h} & \textit{10h} & \textit{0.1h} & \textit{1h} & \textit{10h} \\
      \midrule
      \textit{Baseline$_P$} & 55.17 & \textbf{65.90} & \textbf{71.42} & 9.62 & \textbf{7.31} & \textbf{7.52} \\
      \midrule
      \textit{Most Speakers} & 56.55 & 64.76 & \textbf{71.55} & 9.83 & 8.35 & \textbf{7.74} \\
      \bottomrule
    \end{tabular}
    }
    \label{table:speaker}
\end{table}

\begin{figure}
\centering
    \begin{minipage}[t]{.5\linewidth}
        \includegraphics[width=\linewidth]{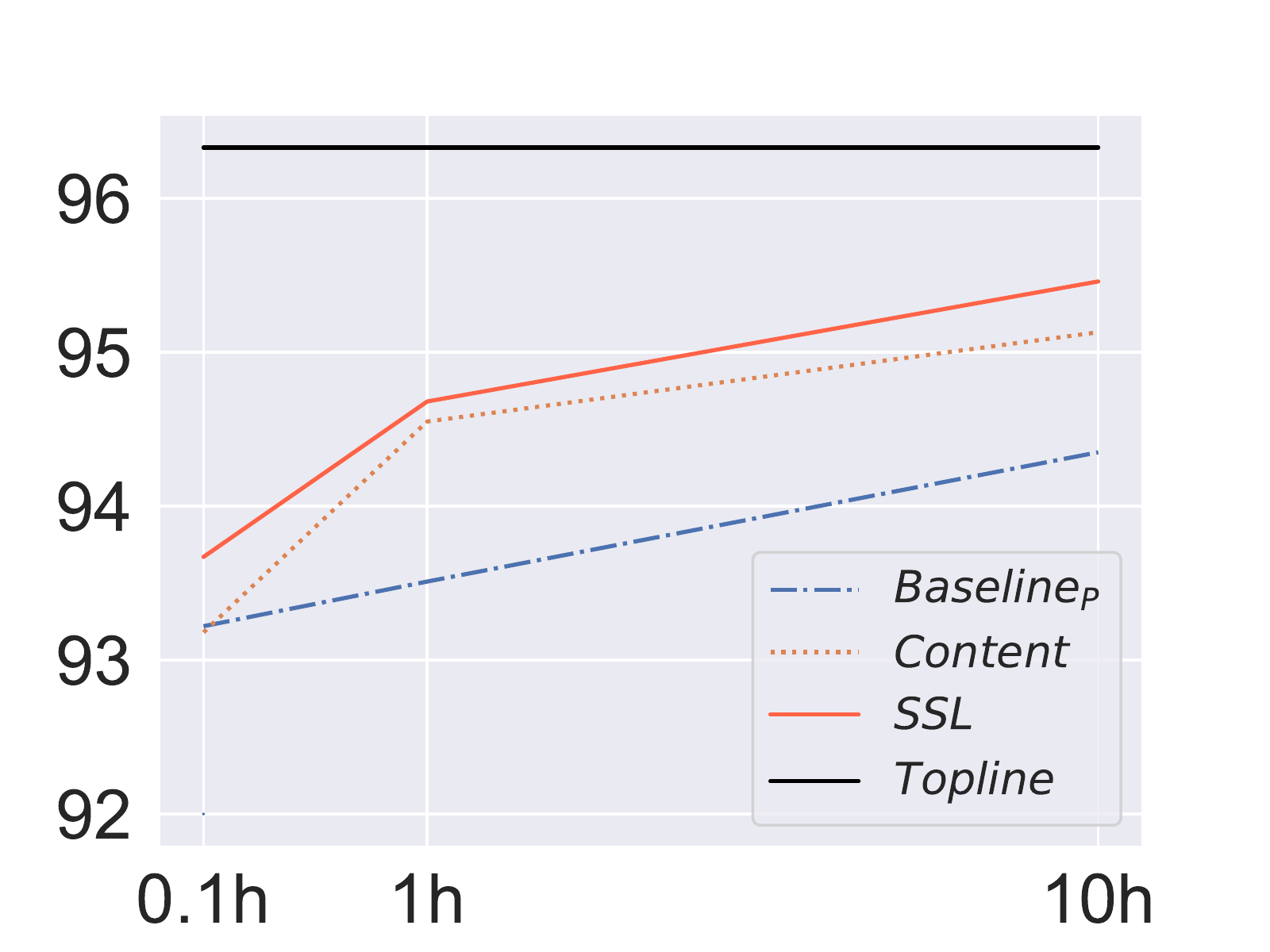}\\%
        \centerline{(a) Accuracy}
    \end{minipage}\hfil
    \begin{minipage}[t]{.5\linewidth}
        \includegraphics[width=\linewidth]{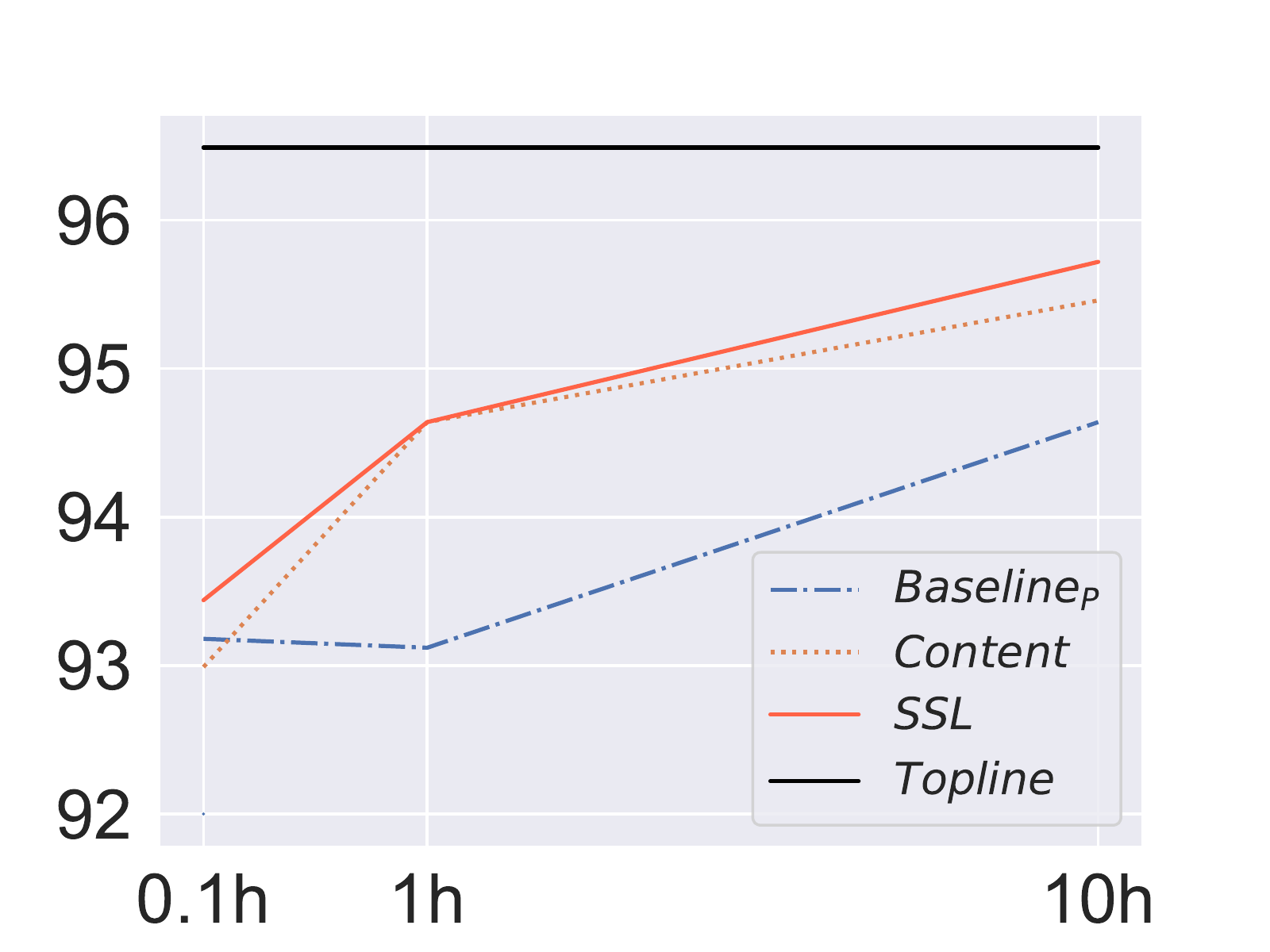}
        \centerline{(b) Agreement}\medskip
    \end{minipage}\hfil
    \caption{Performances of \textit{Baseline$_P$}, \textit{Content}, \textit{SSL}, and \textit{Topline} on KS with victim model WavLM Base+. Agreement refers to the prediction consistency between the victim model and the surrogate model.}
    \label{fig:libri_ks}
\end{figure}

\subsection{Main Results}

 In Table \ref{table:lsds}, \textit{Baseline$_P$} shows an overall-better performance compared to \textit{Baseline$_R$}, especially in the highly low-resource setting, which indicates the effectiveness of pre-training the surrogate model. Note that ``low-resource'' in this paper refers to the limitation on the number of queries.

For the proposed selection methods discussed in Section \ref{sec:method:sm}, \textit{Transcription} has a similar performance compared to \textit{Baseline$_P$}. This indicates that selecting the queries based on speech transcriptions is not effective. From Table \ref{table:lsds} and Figure \ref{fig:libri_ks}, though there’s still room for improvements, our methods have generally outperformed the random selection baseline on KS, IC, and ER, making a step forward in efficient data selection and MEA. Note that similar agreement results as Figure 2(b) can be obtained on IC and ER.$^\text{\ref{note2}}$

From Table \ref{table:lsds}, we observed that all methods and baselines suffer performance degradation in low-resource situations on SD. Therefore, we conducted extensive experiments on another speaker task: speaker identification (SID) \cite{nagrani2020voxceleb}, and provide \textit{Most Speakers}, referring to the most-speaker selection, which samples as many speakers as possible, to tackle speaker tasks. The results are shown in Table \ref{table:speaker}\footnote{\label{note2}Only part of the results are shown due to page limitation.}. However, the performance still does not improve under the low-resource setting.


\subsection{Mismatched Querying Datasets}
\label{exp:dqd}

\begin{table}
    \centering
    \caption{Performances of SD with the victim model HuBERT Base. The querying datasets are denoted in parentheses.}
   \resizebox{\linewidth}{!}{
   \begin{tabular}{l | c c c | c c c}
      \toprule
      & \multicolumn{3}{c|}{\textbf{SD} (Aishell-1)} & \multicolumn{3}{c}{\textbf{SD} (WSJ)} \\
      \cmidrule{2-7}
      \textit{Method} & \textit{0.1h} & \textit{1h} & \textit{10h} & \textit{0.1h} & \textit{1h} & \textit{10h} \\
      \midrule
      \textit{Baseline$_P$} & 9.09 & 8.25 & 7.21 & 8.55 & 8.44 & 7.80 \\
      \midrule
      \textit{SSL} & 8.31 & 7.46 & 6.90 & 9.45 & 7.73 & 7.88 \\
      \midrule
      \textit{Content} & 9.21 & 8.20 & 6.88 & 9.50 & 7.82 & 7.93 \\
      \bottomrule
    \end{tabular}
    }
    \label{table:qd}
\end{table}

\begin{figure}
    \centering
    \begin{minipage}[t]{.5\linewidth}
        \includegraphics[width=\linewidth]{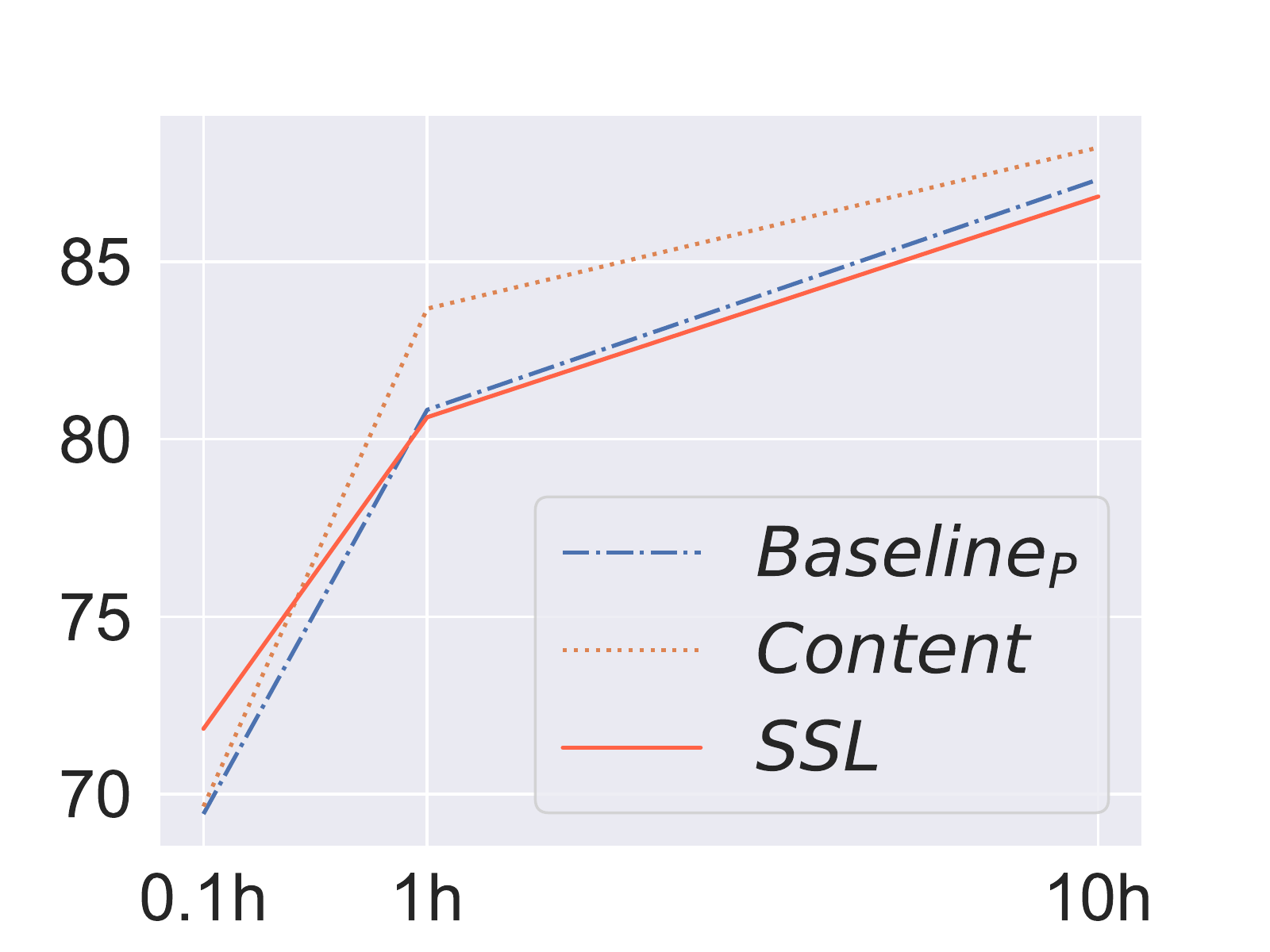}\\%
        \centerline{(a) Aishell-1}\medskip
    \end{minipage}\hfil
    \begin{minipage}[t]{.5\linewidth}
        \includegraphics[width=\linewidth]{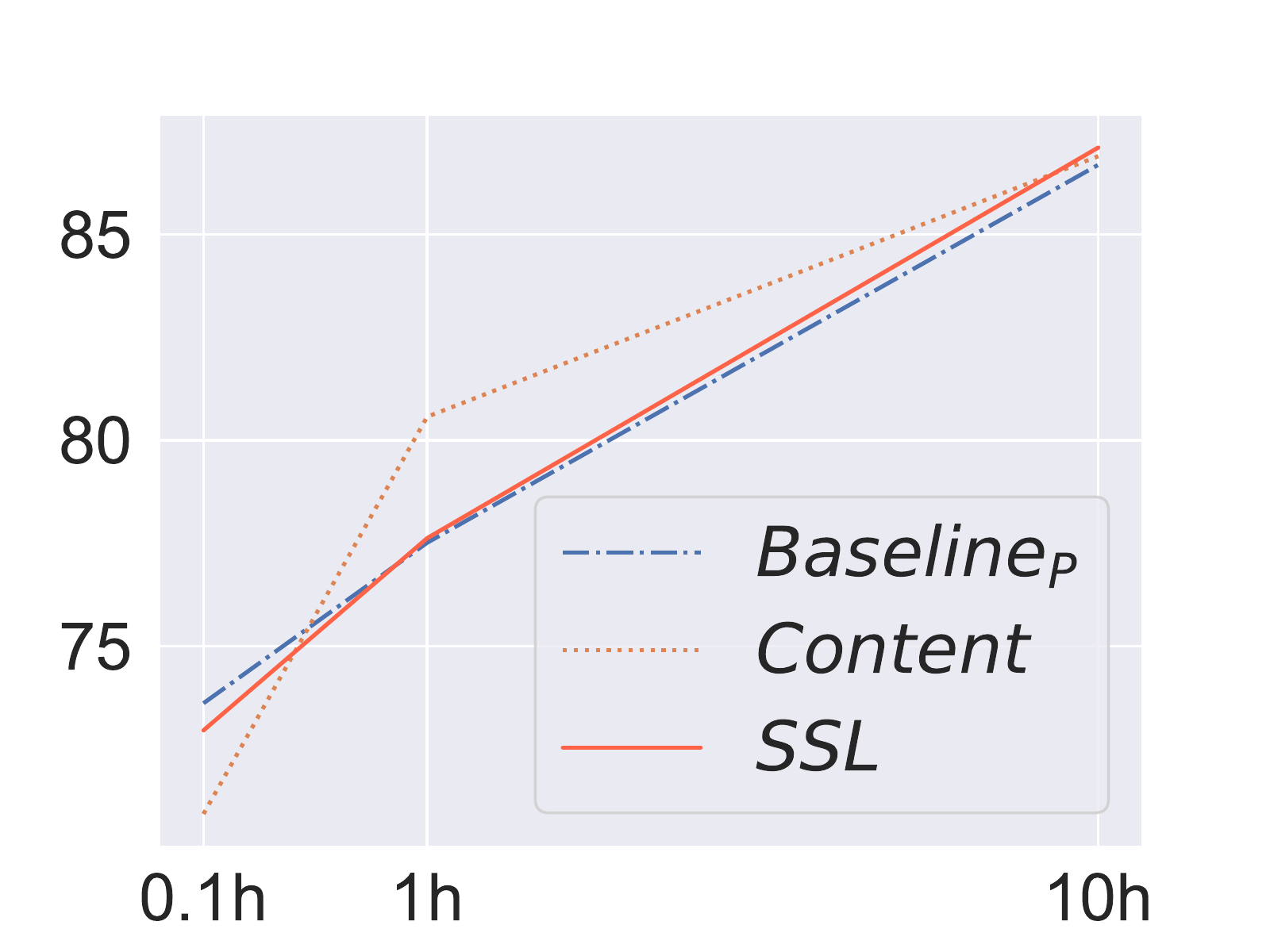}\\%
        \centerline{(b) WSJ}\medskip
    \end{minipage}%
\caption{IC performances of \textit{Content}, \textit{SSL}, and \textit{Baseline$_P$} with querying dataset Aishell-1 and WSJ.}
\label{fig:qd}
\end{figure}
The checkpoints of the pre-trained models are often released while the pre-training datasets are not publicly available due to commercial use, privacy issue, etc. Therefore, we examine WSJ and AISHELL-1 as querying datasets to simulate the situation that the pre-training dataset of \textit{Pre-trained Surrogate} is unknown.
It is worth mentioning that AISHELL-1 is a Chinese corpus, which means the surrogate model's pre-training dataset and our querying dataset are in different languages.

As shown in Figure \ref{fig:qd}, there is no obvious performance difference between our proposed methods and \textit{Baseline$_P$} (the same conclusion can be drawn on KS, ER, and SD). From Table \ref{table:lsds} and Table \ref{table:qd}$^\text{\ref{note2}}$, we observe that extracting the models with the AISHELL-1 corpus achieves a slightly better performance on SD than with the WSJ and the LibriSpeech corpus. On the other hand, extracting the victim model with the LibirSpeech corpus usually outperforms the other two corpora in other tasks, which means that performing model extraction with the different datasets pre-trained on the surrogate model may significantly affect the performance.

\section{Conclusion and Future Works}
\label{sec:conclusion}
This work makes the first attempt to conduct the model extraction attack against SSL speech models. Experimental results on four diverse tasks in SUPERB show that our proposed selection methods outperform the naive random data selection. In the future, we expect to explore a more effective data selection method and find a way to avoid ineffective data selection resulting from the mismatch between pre-training and querying dataset as mentioned in Section \ref{exp:dqd}.

\newpage

\bibliographystyle{IEEEtran}
\bibliography{refs}

\begin{thebibliography}{10}
\providecommand{\url}[1]{#1}
\csname url@samestyle\endcsname
\providecommand{\newblock}{\relax}
\providecommand{\bibinfo}[2]{#2}
\providecommand{\BIBentrySTDinterwordspacing}{\spaceskip=0pt\relax}
\providecommand{\BIBentryALTinterwordstretchfactor}{4}
\providecommand{\BIBentryALTinterwordspacing}{\spaceskip=\fontdimen2\font plus
\BIBentryALTinterwordstretchfactor\fontdimen3\font minus
  \fontdimen4\font\relax}
\providecommand{\BIBforeignlanguage}[2]{{%
\expandafter\ifx\csname l@#1\endcsname\relax
\typeout{** WARNING: IEEEtran.bst: No hyphenation pattern has been}%
\typeout{** loaded for the language `#1'. Using the pattern for}%
\typeout{** the default language instead.}%
\else
\language=\csname l@#1\endcsname
\fi
#2}}
\providecommand{\BIBdecl}{\relax}
\BIBdecl

\bibitem{baevski2020wav2vec}
A.~Baevski, Y.~Zhou, A.~Mohamed, and M.~Auli, ``wav2vec 2.0: A framework for
  self-supervised learning of speech representations,'' \emph{Advances in
  neural information processing systems}, vol.~33, pp. 12\,449--12\,460, 2020.

\bibitem{hsu2021hubert}
W.-N. Hsu, B.~Bolte, Y.-H.~H. Tsai, K.~Lakhotia, R.~Salakhutdinov, and
  A.~Mohamed, ``Hubert: Self-supervised speech representation learning by
  masked prediction of hidden units,'' \emph{IEEE/ACM Transactions on Audio,
  Speech, and Language Processing}, vol.~29, pp. 3451--3460, 2021.

\bibitem{chen2022wavlm}
S.~Chen, C.~Wang, Z.~Chen, Y.~Wu, S.~Liu, Z.~Chen, J.~Li, N.~Kanda,
  T.~Yoshioka, X.~Xiao \emph{et~al.}, ``Wavlm: Large-scale self-supervised
  pre-training for full stack speech processing,'' \emph{IEEE Journal of
  Selected Topics in Signal Processing}, 2022.

\bibitem{yang21c_interspeech}
S.~wen Yang, P.-H. Chi, Y.-S. Chuang, C.-I.~J. Lai, K.~Lakhotia, Y.~Y. Lin,
  A.~T. Liu, J.~Shi, X.~Chang, G.-T. Lin, T.-H. Huang, W.-C. Tseng, K.~tik Lee,
  D.-R. Liu, Z.~Huang, S.~Dong, S.-W. Li, S.~Watanabe, A.~Mohamed, and
  H.~yi~Lee, ``{SUPERB: Speech Processing Universal PERformance Benchmark},''
  in \emph{Proc. Interspeech 2021}, 2021, pp. 1194--1198.

\bibitem{brown2020language}
T.~Brown, B.~Mann, N.~Ryder, M.~Subbiah, J.~D. Kaplan, P.~Dhariwal,
  A.~Neelakantan, P.~Shyam, G.~Sastry, A.~Askell \emph{et~al.}, ``Language
  models are few-shot learners,'' \emph{Advances in neural information
  processing systems}, vol.~33, pp. 1877--1901, 2020.

\bibitem{tramer2016stealing}
F.~Tram{\`e}r, F.~Zhang, A.~Juels, M.~K. Reiter, and T.~Ristenpart, ``Stealing
  machine learning models via prediction $\{$APIs$\}$,'' in \emph{25th USENIX
  security symposium (USENIX Security 16)}, 2016, pp. 601--618.

\bibitem{correia2018copycat}
J.~R. Correia-Silva, R.~F. Berriel, C.~Badue, A.~F. de~Souza, and
  T.~Oliveira-Santos, ``Copycat cnn: Stealing knowledge by persuading
  confession with random non-labeled data,'' in \emph{2018 International Joint
  Conference on Neural Networks (IJCNN)}.\hskip 1em plus 0.5em minus
  0.4em\relax IEEE, 2018, pp. 1--8.

\bibitem{orekondy2019knockoff}
T.~Orekondy, B.~Schiele, and M.~Fritz, ``Knockoff nets: Stealing functionality
  of black-box models,'' in \emph{Proceedings of the IEEE/CVF conference on
  computer vision and pattern recognition}, 2019, pp. 4954--4963.

\bibitem{takemura2020model}
T.~Takemura, N.~Yanai, and T.~Fujiwara, ``Model extraction attacks on recurrent
  neural networks,'' \emph{Journal of Information Processing}, vol.~28, pp.
  1010--1024, 2020.

\bibitem{shen2022model}
Y.~Shen, X.~He, Y.~Han, and Y.~Zhang, ``Model stealing attacks against
  inductive graph neural networks,'' in \emph{2022 IEEE Symposium on Security
  and Privacy (SP)}.\hskip 1em plus 0.5em minus 0.4em\relax IEEE, 2022, pp.
  1175--1192.

\bibitem{wu2022model}
B.~Wu, X.~Yang, S.~Pan, and X.~Yuan, ``Model extraction attacks on graph neural
  networks: Taxonomy and realisation,'' in \emph{Proceedings of the 2022 ACM on
  Asia Conference on Computer and Communications Security}, 2022, pp. 337--350.

\bibitem{wang2018stealing}
B.~Wang and N.~Z. Gong, ``Stealing hyperparameters in machine learning,'' in
  \emph{2018 IEEE symposium on security and privacy (SP)}.\hskip 1em plus 0.5em
  minus 0.4em\relax IEEE, 2018, pp. 36--52.

\bibitem{oh2019towards}
S.~J. Oh, B.~Schiele, and M.~Fritz, ``Towards reverse-engineering black-box
  neural networks,'' in \emph{Explainable AI: Interpreting, Explaining and
  Visualizing Deep Learning}.\hskip 1em plus 0.5em minus 0.4em\relax Springer,
  2019, pp. 121--144.

\bibitem{devlin-etal-2019-bert}
\BIBentryALTinterwordspacing
J.~Devlin, M.-W. Chang, K.~Lee, and K.~Toutanova, ``{BERT}: Pre-training of
  deep bidirectional transformers for language understanding,'' in
  \emph{Proceedings of the 2019 Conference of the North {A}merican Chapter of
  the Association for Computational Linguistics: Human Language Technologies,
  Volume 1 (Long and Short Papers)}.\hskip 1em plus 0.5em minus 0.4em\relax
  Minneapolis, Minnesota: Association for Computational Linguistics, Jun. 2019,
  pp. 4171--4186. [Online]. Available: \url{https://aclanthology.org/N19-1423}
\BIBentrySTDinterwordspacing

\bibitem{yang2019xlnet}
Z.~Yang, Z.~Dai, Y.~Yang, J.~Carbonell, R.~R. Salakhutdinov, and Q.~V. Le,
  ``Xlnet: Generalized autoregressive pretraining for language understanding,''
  \emph{Advances in neural information processing systems}, vol.~32, 2019.

\bibitem{liu2019roberta}
Y.~Liu, M.~Ott, N.~Goyal, J.~Du, M.~Joshi, D.~Chen, O.~Levy, M.~Lewis,
  L.~Zettlemoyer, and V.~Stoyanov, ``Roberta: A robustly optimized bert
  pretraining approach,'' \emph{arXiv preprint arXiv:1907.11692}, 2019.

\bibitem{Krishna2020Thieves}
\BIBentryALTinterwordspacing
K.~Krishna, G.~S. Tomar, A.~P. Parikh, N.~Papernot, and M.~Iyyer, ``Thieves on
  sesame street! model extraction of bert-based apis,'' in \emph{International
  Conference on Learning Representations}, 2020. [Online]. Available:
  \url{https://openreview.net/forum?id=Byl5NREFDr}
\BIBentrySTDinterwordspacing

\bibitem{chen2021killing}
C.~Chen, X.~He, L.~Lyu, and F.~Wu, ``Killing one bird with two stones: Model
  extraction and attribute inference attacks against bert-based apis,''
  \emph{arXiv e-prints}, pp. arXiv--2105, 2021.

\bibitem{he-etal-2021-model}
\BIBentryALTinterwordspacing
X.~He, L.~Lyu, L.~Sun, and Q.~Xu, ``Model extraction and adversarial
  transferability, your {BERT} is vulnerable!'' in \emph{Proceedings of the
  2021 Conference of the North American Chapter of the Association for
  Computational Linguistics: Human Language Technologies}.\hskip 1em plus 0.5em
  minus 0.4em\relax Online: Association for Computational Linguistics, Jun.
  2021, pp. 2006--2012. [Online]. Available:
  \url{https://aclanthology.org/2021.naacl-main.161}
\BIBentrySTDinterwordspacing

\bibitem{zanella2021grey}
S.~Zanella-Beguelin, S.~Tople, A.~Paverd, and B.~K{\"o}pf, ``Grey-box
  extraction of natural language models,'' in \emph{International Conference on
  Machine Learning}.\hskip 1em plus 0.5em minus 0.4em\relax PMLR, 2021, pp.
  12\,278--12\,286.

\bibitem{eldar1997farthest}
Y.~Eldar, M.~Lindenbaum, M.~Porat, and Y.~Y. Zeevi, ``The farthest point
  strategy for progressive image sampling,'' \emph{IEEE Transactions on Image
  Processing}, vol.~6, no.~9, pp. 1305--1315, 1997.

\bibitem{chang2022distilhubert}
H.-J. Chang, S.-w. Yang, and H.-y. Lee, ``Distilhubert: Speech representation
  learning by layer-wise distillation of hidden-unit bert,'' in \emph{ICASSP
  2022-2022 IEEE International Conference on Acoustics, Speech and Signal
  Processing (ICASSP)}.\hskip 1em plus 0.5em minus 0.4em\relax IEEE, 2022, pp.
  7087--7091.

\bibitem{ott2019fairseq}
M.~Ott, S.~Edunov, A.~Baevski, A.~Fan, S.~Gross, N.~Ng, D.~Grangier, and
  M.~Auli, ``fairseq: A fast, extensible toolkit for sequence modeling,'' in
  \emph{Proceedings of NAACL-HLT 2019: Demonstrations}, 2019.

\bibitem{panayotov2015librispeech}
V.~Panayotov, G.~Chen, D.~Povey, and S.~Khudanpur, ``Librispeech: an asr corpus
  based on public domain audio books,'' in \emph{2015 IEEE international
  conference on acoustics, speech and signal processing (ICASSP)}.\hskip 1em
  plus 0.5em minus 0.4em\relax IEEE, 2015, pp. 5206--5210.

\bibitem{paul1992design}
D.~B. Paul and J.~Baker, ``The design for the wall street journal-based csr
  corpus,'' in \emph{Speech and Natural Language: Proceedings of a Workshop
  Held at Harriman, New York, February 23-26, 1992}, 1992.

\bibitem{bu2017aishell}
H.~Bu, J.~Du, X.~Na, B.~Wu, and H.~Zheng, ``Aishell-1: An open-source mandarin
  speech corpus and a speech recognition baseline,'' in \emph{2017 20th
  conference of the oriental chapter of the international coordinating
  committee on speech databases and speech I/O systems and assessment
  (O-COCOSDA)}.\hskip 1em plus 0.5em minus 0.4em\relax IEEE, 2017, pp. 1--5.

\bibitem{lloyd1982least}
S.~Lloyd, ``Least squares quantization in pcm,'' \emph{IEEE transactions on
  information theory}, vol.~28, no.~2, pp. 129--137, 1982.

\bibitem{speechcommands}
P.~Warden, ``Speech commands: A public dataset for single-word speech
  recognition.'' \emph{Dataset available from
  http://download.tensorflow.org/data/speech\_commands\_v0.01.tar.gz}, 2017.

\bibitem{Lugosch2019}
\BIBentryALTinterwordspacing
L.~Lugosch, M.~Ravanelli, P.~Ignoto, V.~S. Tomar, and Y.~Bengio, ``{Speech
  Model Pre-Training for End-to-End Spoken Language Understanding},'' in
  \emph{Proc. Interspeech 2019}, 2019, pp. 814--818. [Online]. Available:
  \url{http://dx.doi.org/10.21437/Interspeech.2019-2396}
\BIBentrySTDinterwordspacing

\bibitem{busso2008iemocap}
C.~Busso, M.~Bulut, C.-C. Lee, A.~Kazemzadeh, E.~Mower, S.~Kim, J.~N. Chang,
  S.~Lee, and S.~S. Narayanan, ``Iemocap: Interactive emotional dyadic motion
  capture database,'' \emph{Language resources and evaluation}, vol.~42, pp.
  335--359, 2008.

\bibitem{cosentino2020librimix}
J.~Cosentino, M.~Pariente, S.~Cornell, A.~Deleforge, and E.~Vincent,
  ``Librimix: An open-source dataset for generalizable speech separation,''
  \emph{arXiv preprint arXiv:2005.11262}, 2020.

\bibitem{nagrani2020voxceleb}
A.~Nagrani, J.~S. Chung, W.~Xie, and A.~Zisserman, ``Voxceleb: Large-scale
  speaker verification in the wild,'' \emph{Computer Speech \& Language},
  vol.~60, p. 101027, 2020.

\end{thebibliography}

\end{document}